Comparing the efficacy of fixed effect and MAIHDA models in predicting outcomes for intersectional social strata

Ben Van Dusen – Iowa State University
Heidi Cian – Maine Mathematics and Science Alliance
Jayson Nissen – Nissen Education Research and Design
Lucy Arellano – Texas Technology University
Adrienne Woods - SRI International

Originating in 1980s legal scholarship, the analytical lens of *intersectionality* brought attention to "the vexed dynamics of difference and the solidarities of sameness in the context of antidiscrimination and social movement politics" (Cho et al. 2013, p. 787). In this way, the notion of intersectionality brought to light the "sameness" of experiences felt by a designated demographic group (e.g., "women") but, uniquely, also the essential differences in experiences that resulted from overlapping identities (e.g., "white woman" and "Black woman"). Since Crenshaw (1990) translated the term from activist communities into formal scholarship, qualitative investigators have used it to consider interactions between individuals' multi-faceted social identities and the power structures of their contexts to which components of their intersecting identities are subjected (Collins 2019). While theorized initially to consider the experiences of Black women in the legal system, intersectionality research has evolved to include a range of intersectional identities, including sex/gender, age, social class, race, sexual orientation/identity, disability, and immigration status across fields of study (Harris and Patton 2019; Mena and Bolte 2019).

Salem (2018) described intersectionality as a theory that travels across time, place, and space. As intersectionality has made this conceptual journey, it has been adapted and used differently across contexts and communities. The versatile nature of intersectionality is captured by Tefera and colleagues (2018), calling it "aspirational." Carbado and colleagues (2013, p. 304) go further to say that "[intersectionaly] is never done, nor exhausted by its prior articulations or movements; it is always already an analysis-in-progress."

With the emergence of critical quantitative theories (Tabron and Thomas 2023), the theory of intersectionality has begun to move into quantitative spaces. This transformation has led some *quantitative* researchers to explore ways to take up theoretical constructs typically associated with qualitative critical investigations–like intersectionality–using quantitative methods (Wofford and Winkler 2022). The ad hoc nature of this transformation and existing methods have limited quantitative researchers' practical and theoretical ability to enact intersectional research. This limitation can be seen in Bauer and colleagues' (2021) review of intersectionality quantitative research in which they find that existing work largely lacks evidence of theoretical alignment. It can be difficult to imagine something when we cannot create it. Creating new methods can unlock the development of theory (Alvesson and Kärreman, 2007).

Research points to a need to investigate and address inequities in science, technology, engineering, and mathematics (STEM) higher education using intersectionality. Extensive research details the systemic exclusion of individuals marginalized by racist, sexist, classist, and ableist standards of STEM participation (e.g., Carter et al. 2019; Ireland et al. 2018; Kim et al. 2018) and identifies course and institutional structures that educators can change to mitigate these inequities (e.g., Denton and Borrego 2021; Diaz-Eaton et al. 2022; Jackson et al. 2021).

While many qualitative studies use intersectionality to inform the experiences and marginalization of Black women (Nash, 2008) and women of color in STEM broadly or in specific STEM fields, quantitative research seldom employs intersectional analyses (cf. Byars-Winston and Rodgers, 2019). This existing literature suggests that the discursive structures that define STEM participation similarly require negotiation and cultural "trade-offs" for individuals identifying with groups marginalized by white, masculine, and ableist standards of participation. However, the complexity of this negotiation escalates when considering intersecting ways of identifying, particularly if individuals attempt to bring their whole selves to the STEM space. For instance, Black women and women of color are often ignored and avoided by peers and faculty members, discouraged from pursuing STEM degrees by faculty members, excluded from insider know-how needed to succeed in their education, and treated as 'too smart' by their peers and family (Dortch and Patel, 2017; Johnson, 2001; Johnson, 2006; McGee and Bentley, 2017; McPherson, 2017; Ong, 2005; Seymour, 2000; Hyatar-Adams et al., 2019).

Quantitative equity research in STEM higher education tends to aggregate marginalized groups (e.g., underrepresented minorities, URMs) and to separate the analysis across oppressive forces (e.g., racism, sexism, class oppression). Theory seldom drives these analytical decisions; instead, they are made to align with institutional policies or practices, e.g., the National Science Foundation's use of URM, in the pursuit of statistically significant results (see critiques in Wasserstein and Lazar). These aggregations obscure the inequities experienced by marginalized groups (Shafer et al. 2021). By disaggregating the experiences wrought by these oppressive forces individually, applying intersectionality to quantitative equity research in STEM higher education may further the understanding of the sources of these inequities and the transformations to educational practices and policies that create equitable STEM courses (e.g., Van Dusen and Nissen 2020a; 2020b). However, doing so requires analytical techniques that can account for the intersections of social identities and the power structures and practices that vary across courses and STEM disciplines.

In reviewing the use of intersectionality in feminist literature, McCall (2005) identified three approaches applied to intersectional studies: anticategorical, intracategorical, and intercategorical. Quantitative research inherently applies categories and thus falls into the intra or intercategorical approaches (Bauer et al. 2021), with intercategorical approaches being the most common (Bauer et al. 2021). The fixed effects approach is the most used method for creating models that examine intersectional outcomes (Evans, 2018). In a fixed effects approach, researchers create models with main effects for each axis of intersectional social strata and include all of the possible interaction terms in the model (Evans et al. 2020). Intersectional social strata are the provisionally adopted analytical categories (e.g., race, gender, and first-generation designation) used to document inequalities. While this method creates predicted outcomes for each intersectional social strata that are more nuanced than simply adding the main effects, it becomes cumbersome as more strata are added. The number of interaction terms grows exponentially. As such, as the number of terms included in the model grows, collecting datasets with sufficient statistical power can be prohibitive.

Researchers have proposed *a multilevel analysis of individual heterogeneity and discriminatory accuracy* (MAIHDA; Evans 2015; Evans et al. 2018; Merlo 2018; Evans 2019) as a theoretically engaged approach designed to address some of the shortcomings of prior methods. To improve model predictions across strata, MAIHDA nests individuals within their social identities (Figure 1; Evans et al. 2020; Keller et al. 2023). By combining the main effects with the variance terms for each strata, MAIHDA can create more accurate predictions than adding the primary terms without including interaction terms. MAIHDA offers several potential advantages over fixed effects models. First, MAIHDA aligns with an intersectional perspective by including all strata in a model, regardless of their sample size. Second, by reducing the number of terms in the model, MAIHDA reduces the statistical power requirements. Third, MAIHDA seeks to increase the accuracy of predictions by drawing on the shrinkage, or partial pooling, that nesting within each aspect of a strata provides in a multilevel model (Raudenbush and Bryk, 1986). Shrinkage allows multilevel models to make predictions for each strata that are informed by the predictions made for other strata. Shrinkage benefits are likely to be strongest for small-N strata where the small numbers limit the abilities of intersectional models to disaggregate outcomes.

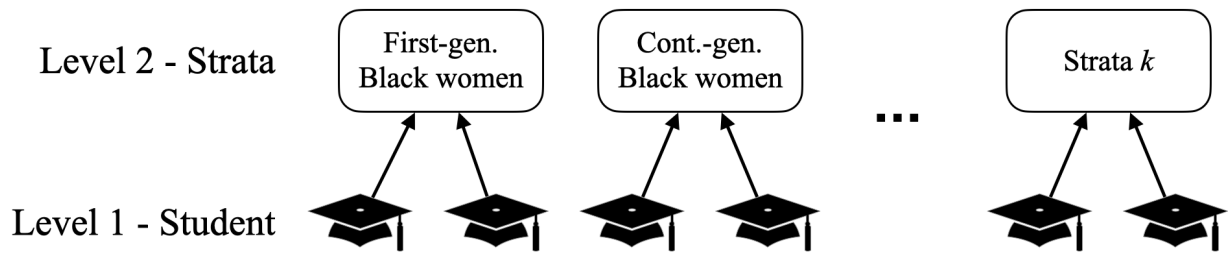

**Figure 1.** MAIHDA nesting students within *k* strata.

While MAIHDA has many promising features for building models that account for intersecting axes of social identities, more simulation studies are needed to demonstrate the method's efficacy empirically. Existing efforts have used simplified simulations (e.g., Lizotte et al. 2020; Evans et al. 2020; Bell et al. 2019) or simulations mirroring health outcomes (e.g., Mahendran et al. 2022a; 2022b). No simulation study to date shows how MAIHDA impacts the accuracy of predicted strata outcomes in education and how this improvement in accuracy might enable examining strata with smaller sample sizes. Nor have these simulation studies addressed the nested nature of data that commonly exists in education studies: measurements within students within courses within schools.

In this simulation study, we address this need by comparing the efficacy of using MAIHDA versus fixed effects models to examine outcomes for intersectional social strata. To assess model performance in a real-world scenario likely to be of interest to equity researchers, we grounded our simulation in data from a widely used assessment (the Force Concept Inventory; Hestenes, Wells, and Swackhamer 1992) in a field with well-documented inequities (i.e., physics; Brewe and Sawtelle, 2016). Our simulations are based on real-world strata performances and participation rates. The simulations nest students in courses, include course-level variation, and vary in size to represent the sample sizes common in science equity publications (Van Dusen and Nissen 2022). While we simplified course-level variation, we accounted for students nested in courses, as this nesting is common and can impact predicted student outcomes (Van Dusen and Nissen, 2019).

# Research Questions

To understand the efficacy of MAIHDA for examining outcomes for intersectional social strata, we asked the following three research questions:
1. What are typical strata representation rates and outcomes on physics research-based assessments?

2. To what extent do MAIHDA models create more accurate predicted strata outcomes than fixed effects models?
3. To what extent do MAIHDA models allow the modeling of smaller strata sample sizes?

# Definitions

We present a few definitions of terms in Table 1 that we use throughout the manuscript.

*Table 1*. Definition of Terms.

| Term | Definition |
|---|---|
| Accuracy | The difference between the estimated score and the true score. The true error quantifies it. |
| Bayesian statistics | A statistical system based on calculating the probability of a hypothesis given the data, or P(hypothesis\|data). |
| Cross-classified multilevel model | An extension to multilevel models that allows for higher-level units to share a level. For example, a student (level 1) can be nested in both an advisor (level 2a) and course (level 2b). |
| Estimated score | A regression model's predicted score for a strata. |
| First-generation college student | A student whose parent(s) or guardian(s) do not have a bachelor's degree. We use first-generation college designation as an indicator of socioeconomic status. |
| Frequentist statistics | A statistical system based on calculating the probability of the data given a hypothesis, or P(data\|hypothesis). |
| Intersectional social strata | Provisionally adopted analytical categories used to document inequalities (McCall 2005). We focus on strata at the intersection of race, socioeconomic status, and gender. |
| Intersectionality | Analytic lens that views social categories that are subjected to power dynamics (e.g., race, class, and gender) not as discrete and mutually exclusive entities but that they build on each other and work together (Collins and Bilge 2020). |
| Proportional change in variance (PCV) | The total between-stratum variation explained by including additive main effects. PCV is calculated by dividing the difference between the stratum-level variances obtained in the null and additive effects models by the stratum-level variance obtained in the null model. |

| Shrinkage | A statistical feature of multilevel models by which the predicted outcomes for a strata is informed by data from other strata (Raudenbush and Bryk, 1986). |
|---|---|
| True error | The difference between the estimated score and the true score. We calculate true error as the estimate score minus the true score. Negative values are classified as underestimates and positive values are overestimates. |
| True score | A group's true score is a value with no error that is set by the simulation. Observed scores equal the true scores plus random error. We set the true score for each stratum equal to their average score in our real-world data. |
| Variance partition coefficient (VPC) | The portion of the variance attributed to a level in a multilevel model. VCP is calculated by dividing the between-strata variance by the total variance. |

# Intersectionality

Since its introduction by Crenshaw (1989), the concept of intersectionality has been adapted and adopted by scholars across many fields (McCall 2005). Choo and Ferree (2010) and Collins (2015) provide extensive guidelines for applying intersectionality in sociology. Collins advocates for developing intersectionality as a critical social theory to understand and change the existing social order. To use intersectionality in pursuit of these aims, Collins details four principles of power and social structures of which researchers should be aware: (1) social strata (e.g., race, class, and gender) act as markers of power that are interdependent and mutually constructed; (2) power relations across the intersections of these power markers create complex, interdependent inequalities; (3) individuals' and groups' locations within these intersecting power relations shapes their experiences and perspectives; and (4) solving social problems requires intersectional analyses specific to the context of the social problem.

These principles direct quantitative research to build meaningful intersectional models. We use "meaningful" here to differentiate from "significant" in statistical interpretations where statistical significance is used to specify models. Modeling decisions based on statistical significance criteria can fail to identify educationally meaningful differences (Wasserstein et al. 2015) and produce misleading model results (Van Dusen and Nissen 2023). The principles also require that models account for power structures beyond the social strata of individuals in the models, such as the instructional methods used in the courses or the characteristics of the schools.

Choo and Ferree (2010) distinguish between group, process, and system-centered intersectional research. Group-centered research centers the voices of multiple marginalized groups and would fall into McCall's (2005) anti-categorical or intracategorical paradigms. Anticategorical complexity deconstructs analytical categories as simplifications that reproduce inequalities by upholding differences. Intracategorical complexity uses categories strategically by interrogating the production of boundaries while focusing on specific intersectional strata to highlight their lived experience. Research on intersectionality as a process emphasizes power relations and accounts for the multiplicative oppression or privilege at different points of intersection. The emphasis on power relations leads to comparisons that align with McCall's intracategorical, looking at an intersectional group across different social locations, or intercategorical research. Intercategorical complexity requires the temporary use of existing strata to document inequality and the way inequality changes over time or across social locations. System-centered research pushes analyses away from associating inequality with a single system of oppression (e.g., family and gender, education, and class (Willis 1981)). Instead, it looks at the complex interactions across racism, sexism, classism, and other forms of oppression. This research often falls into McCall's intercategorical complexity because it makes comparisons across social strata. System-centered research may also investigate different social locations (e.g., STEM disciplines, institutional characteristics, instructor pedagogy).

Cho, Crenshaw, and McCall (2013) identified the need for developing a field of Intersectionality Studies. Collins (2019) similarly argues for a robust intersectional community that incorporates people who theorize from the bottom up and from the top down – a community where many perspectives, methods, and disciplines act to address the conceptual blindspots within any one group that can limit theoretical advancements. Both argue for intersectionality as a critical practice that engages political projects to transform current unjust systems. Quantitative intersectional studies can further both efforts to create opportunities for populations marginalized by exclusionary expectations of STEM participation, such as through the development and use of effective pedagogies, and to transform the status quo, for example, of what it means to participate in STEM.

# Positionality

Including a positionality statement is common for work that takes a critical perspective. Positionality statements acknowledge a researcher's strata's role in shaping their investigation. Below are the authors' positionality statements.

{Anonymized Author 1}: I identify as a continuing-generation White cisgender, heterosexual man with a hearing impairment. I was raised in low-income households but now earn an upper-middle-class income. People with similar privileges have created and maintained our society's unjust power structures. I believe it is the obligation of those with the privilege to dismantle oppressive systems. My privilege, however, limits my perspective on the lived experiences of marginalized individuals.

{Anonymized Author 2}: I am drawn to studying intersectionality and the use of social groups in research to understand how learners of all ages can experience STEM in ways that affirm how they identify with their bodies, communities, and histories. I am particularly interested in how intersectional research can consider group membership beyond the traditional demographic categories of gender and race and enlighten understanding of how individuals experience STEM opportunities in ways that respect or reject their politics and cultural wealth. I identify as a White cisgender heterosexual woman with centrist political views and a rural, lower socioeconomic upbringing.

{Anonymized Author 3}: Identifying as a White, cisgendered, heterosexual man provides me with opportunities denied to others in American society. My experience growing up poor and serving in the all-male submarine service motivated me to reflect on and work to dismantle oppressive power structures in science.

{Anonymized Author 4}: As a Xicana from East L.A., raised in a low-income household, the first in my family to be born in the U.S., the first to attend college, moving 2,500 miles away from home to enter an undergraduate STEM major at a prestigious university, I utilize intersectionality and quantitative methods to understand my own journey through higher education. Now, utilizing my privilege as a cisgender, heterosexual professor at an R-1 university, I endeavor to spotlight the multiple systems of oppression endured by minoritized student groups in order to transform postsecondary institutions.

{Anonymized Author 5}: My work is in service of expanding educational and societal opportunities for vulnerable or historically underserved children, particularly children with disabilities. I typically analyze large, longitudinal datasets using advanced quantitative methods to better understand whether, how, and to what extent early special education intervention and services may remediate or exacerbate existing educational inequities for children with disabilities. I acknowledge my privilege as a non-disabled White cisgender heterosexual woman from a middle-class background. My perspectives on the experiences of members of the disability community are thus limited but informed by a close relative's experiences within the U.S. education system.

# Background/literature review

## Intersectionality in STEM Higher Education

Most intersectional research in STEM higher education has used qualitative methods to investigate the double-bind that racism and sexism pose to women of color in STEM fields. These studies often focus on physics, engineering, and computer science where the marginalization of women of color illuminates the hidden cultural norms and expectations of the exclusive cultures: outward demonstrations of mastery, competitiveness, self-reliance, individualism, bureaucratic gatekeeping, silence about abuses, and enforcement of belonging based on gender, race, and class (Ong 2023, McGee 2023; Cochran, Boveda and Prescod-Weinstein 2020; Womack et al. 2023). In physics within the United States, the culture of no culture, a shared belief that physics lies above and beyond the disorder of human space and time, hides the shared cultural values (Traweek 2009; Ong 2023). These hidden cultural values and expectations especially exclude students, faculty, and professionals from multiply marginalized groups (Ong 2023; McGee 2023). They also enable White faculty to state a commitment to inclusion while engaging in color-evasive racism that dismisses the racism and sexism women of color face (Robertson et al. 2023; King, Russo-Tait and Andrews 2022; Dancy and Hodari 2023)

Few quantitative studies in STEM higher education build intersectional models. Those who have used interaction terms (Van Dusen and Nissen 2023) to build intersectional models. The common use of p-value cutoffs to exclude interaction terms and small sample sizes results in many studies not building intersectional models (Van Dusen and Nissen 2023; Stewart et al. 2021). These studies have found meaningful differences in content knowledge and beliefs before instruction (Van Dusen et al. 2021; Nissen, Her Many Horses and Van Dusen 2021; Van Dusen and Nissen 2020b), opportunities and performance in AP physics and chemistry courses (Krakehl and Kelly 2021; Palermo, Kelly and Krakehl 2022), and inequities in course failure rates (Van Dusen and Nissen 2020a) that represent the educational debt American society owes to Black, Brown, Indigenous, and poor students (Ladson-Billings 2006) and women in STEM. Chemistry courses using student-centered pedagogies reduce these differences in content knowledge and partially repay these educational debts (Van Dusen et al. 2021). Physics courses using student-centered and lecture-based instruction maintain differences in content knowledge and beliefs that perpetuate these educational debts and thereby exclude students from multiply marginalized groups (Nissen, Her Many Horses and Van Dusen 2021; Van Dusen and Nissen 2020b).

## Approaches to Modeling Intersectionality

### Fixed effects approach

Bauer (2014) notes that intersectionality *theory* has struggled to match with intersectional quantitative *methods* despite the potential for quantitative study and intersectionality theory to advance one another mutually. Awareness of the necessity to consider intersectional strata in analysis has especially been noted in health research, where researchers recognize the complexity of health-related inequities and the necessity for quantitative tools to explore the causes and consequences of these inequities (ibid.). However, the effects of quantitatively applying intersectionality in education have also been illuminating. Quantitative intersectionality research has examined the confluence of intersecting social identities (i.e., race, gender, and socioeconomic status) and their connected social power structures in academic outcomes (Riegle-Crumb and Grodsky 2010), disciplinary action (Morris and Perry 2017), and college savings (Quadlin and Conwell 2021). Intersectionality lenses can also be applied to explore the effects of less-considered social strata as an intersectional factor of outcomes, such as studying the effects of body size, race, and sex on academic outcomes (Branigan 2017).

Researchers have used various analytical tools to consider intersectionality in their quantitative research. In a review of quantitative intersectional research in health studies, Mena and Bolte (2019) observed that intersections are typically studied in regression equations by including interaction terms or conducting stratified analysis –a finding supported by Bauer et al. (2021) across disciplines. These reviews show that exploring intersectionality in quantitative models has typically been performed by creating regression models with interaction terms for each axis of an individual's strata (e.g., gender x race) (Evans 2019). For instance, Conwell (2021) used cubic regression with intersectional terms for race and income strata to examine the effects of race and parental income on children's mathematics scores. Riegle-Crumb and Grodsky (2010) considered a variety of school and student factors to conduct multivariate regression analysis that considered interaction terms, such as "Hispanic x Family Income," "Hispanic x Parental Education Level," and "Hispanic x Percentage Minority in School."

Researchers have also recognized that these intersectional approaches can be informative in addressing questions involving nested data and data that evaluates intervention treatments. For instance, Morris and Perry (2017) studied the interaction of race and gender on office referrals. The authors conducted a series of models including an interaction term for race/ethnicity and gender and using a three-level model, with observations (level 1) nested within students (level 2) nested within schools (level 3).

### Problems with fixed effects approaches

While the approach of using interaction terms introduced above allows for the analysis of intersectional social strata, it quickly becomes unwieldy as strata numbers increase and may limit the precision of the model (Evans et al. 2018). Additionally, such studies often involve large national data sets of over 1,000 students (Bauer et al. 2021), limiting their practical use in many research contexts and purposes where obtaining these response numbers is not practical. The sample size limitation is consequential to authentic questions of intersectionality across programs of study because of the need for disaggregation and sufficient representation of strata, as disaggregation requires having enough responses from each strata in the sample. Further, researchers interested in examining intersectional experiences across treatment groups or within nested structures must combine various techniques, potentially increasing model errors. For instance, in their study of mathematics achievement, as predicted by racial/ethnic disparities and advanced mathematics coursework, Riegle-Crumb and Grodsky (2010) conducted separate multilevel multivariate regression analyses of students according to the category of math coursework (i.e., advanced or not advanced), using intersectional terms. By splitting the data across models, the statistical power is decreased, the model uncertainties are increased, and inequities between strata are more challenging to identify.

## MAIHDA

The points above suggest that fixed effects approaches for studying intersectionality using interaction terms are plagued by logistical and theoretical problems. Scott and Siltanen (2017) used a feminist framing of intersectionality to explore the degree to which this commonly-used form of intersectional study–multiple regression–aligns with what is "conceptually central" to the analysis of intersectionality: attention to context, a focus on the need for open inquiry of inequity without preconceived expectations, and recognition of the multifaceted and multilayered nature of inequities. They found traditionally-used regression analyses with intersectional terms ill-suited to this task and saw the most promise in approaches that considered individual-level data as nested within contexts. While such recommendations have been slow to gain traction in practice, some promising preliminary outcomes using simulated data suggest that a *multilevel analysis of individual heterogeneity and discriminatory accuracy* (MAIHDA; Merlo 2018) may provide a practical solution. MAIHDA uses

> Hierarchical and multilevel models to study large numbers of interactions and intersectional identities while partitioning the total variance between two levels–the *between-strata* (or between-category) level and the *within-strata* (or within-category) level (Evans et al., 2018 p. 64).

Using MAIHDA to develop intersectional models was first proposed in Evans' dissertation (2015). In a continuation of that work, Evans *et al*. (2018) demonstrated the use of MAIHDA on models of health inequities. They found that MAIHDA's multilevel approach is superior to fixed effects approaches using intersectional terms in several vital ways. They note that MAIHDA (1) is easier to manage as intersectional terms are added, (2) may provide more accurate estimations for strata with smaller samples, (3) allows for the considerations of intersectional effects of those who identify with multiply marginalized strata as well as those who occupy some privileged and some disadvantaged categories, (4) provides statistics that are more informative than the statistical significance of interaction terms, and (5) more readily allows the study of within-group variance. Further, the multiple-level and cross-classified opportunities inherent in multilevel models may layer experimental and quasi-experimental groupings upon within-strata levels. Evans *et al.* (2018) empirically demonstrated MAIHDA's superior use of statistical power by simulating data with the number of strata ranging from 4-384 and compared the Bayesian information criterion (BIC; Vrieze 2012) scores for the associated fixed effect and MAIHDA models. They found that the fixed effect BIC scores grew exponentially with adding new strata, while the MAIHDA BIC scores were nearly static.

Independently, Jones *et al*. (2016) examined an analogous technique to MAIHDA. Rather than using the statistical technique to model intersectional outcomes, they modeled voting rates using fixed effects for primary terms and random effects to account for interaction terms. They found that using random effects improved the level of detail in the model and protected researchers from over-interpreting their data.

Since these initial publications, simulation studies have examined different aspects of MAIHDA's efficacy and utility. Below, we describe these studies chronologically to illustrate the evolution of the MAIHDA conversation. Collectively, this work has evaluated MAIHDA in terms of its performance compared to alternatives in a) reducing false positives and b) yielding accurate models with varying sample sizes.

Bell *et al*. (2019) performed a simulation study examining false-positive rates for intersectional terms in fixed effect and MAIHDA models. They found that MAIHDA models produced fewer false positives than the saturated fixed effect models.

Lizotte *et al.* (2020) demonstrated that when given large sample sizes (100,000) and large strata populations (3,125), the MAIHDA models and the fixed effect models were both highly accurate. They also proposed that Evans (2018) had misinterpreted MAIHDA model terms. Specifically, they claimed that the fixed effects had been mistaken for the grand means, and the residual terms had been mistaken for the intersectional effects. Evans *et al*. (2020) rebutted the claims about their misinterpreting terms. Using a simulation study, they demonstrated that, like with all multilevel models, the fixed effects in MAIDHA models are precision-weighted grand means, which, under some conditions, will be equal to the grand means.

Mahendran *et al.* (2022a; 2022b) produced a pair of simulation studies comparing the efficacy of seven different modeling techniques (i.e., cross-classification, regression with interactions, MAIHDA, and decision trees [CART, CTree, CHAID, and random forest]) to examine intersectional health outcomes. First, they modeled binary outcomes using logistic regression. They found that, while some methods could match MAIHDA's efficacy under specific conditions, no method outperformed MAIHDA in estimation accuracy or sensitivity analysis under any condition (Mahendran *et al*. 2022a). In the subsequent study (Mahendran *et al*. 2022b), they modeled continuous outcomes. They concluded that, while random forest was generally the most robust method, MAIHDA was still recommended for all sample sizes. Both studies diverged from the traditional use of Bayesian models with MAIHDA in favor of frequentist models. They claimed that their analysis showed no meaningful differences between the two methods and preferred the frequentist models because they could be estimated more quickly.

While the above research primarily assessed health data, MAIHDA has only recently begun to be used in education research. Keller *et al*. (2023) systematically reviewed MAIHDA's use and applied it to an educational context (i.e., reading achievement scores). As with the prior investigations, they concluded that MAIHDA offers better scalability for higher dimensions, model parsimony, and precision-weighted estimates of strata with small Ns compared to fixed effect models. They also note, however, several potential disadvantages. One, it can be difficult in educational contexts to collect large enough sample sizes to model the range of students' intersectional strata. Additionally, they point to the need for future research to explore how MAIHDA can account for the nested nature of student data (e.g., students in courses and schools). While Evans (2019) did a robustness check on whether their MAIHDA model would be improved by nesting individuals in school as well as social strata (creating a cross-classified multilevel model), no existing studies presented findings from cross-classified multilevel MAIHDA models.

In this simulation study, we expand on the literature by creating a realistic model of science student outcomes–in which strata composition varies to mirror likely data sets–and comparing the accuracy of MAIHDA against fixed effects models of varying sample sizes. This study has several novel aspects that further examine the efficacy of using MAIDHA in educational contexts. We improved the sophistication of the simulated data from prior studies by building it off of large-scale, real-world education data. We used cross-classified multilevel models that nest students in courses and social strata, a critical feature of large-scale educational studies (Raudenbush and Bryk, 1986; DiPrete and Forristal, 1994; Niehaus et al. 2014; Van Dusen and Nissen 2019). We also examine how MAIHDA's improved accuracy can allow researchers to include more strata in their models without sacrificing the accuracy of their predictions.

# Methods

To compare the accuracy and precision of fixed effects and MAIHDA models across three sample sizes, our analysis included four steps: 1) data collection, 2) data simulation, 3) modeling, and 4) model comparison (Figure 2).

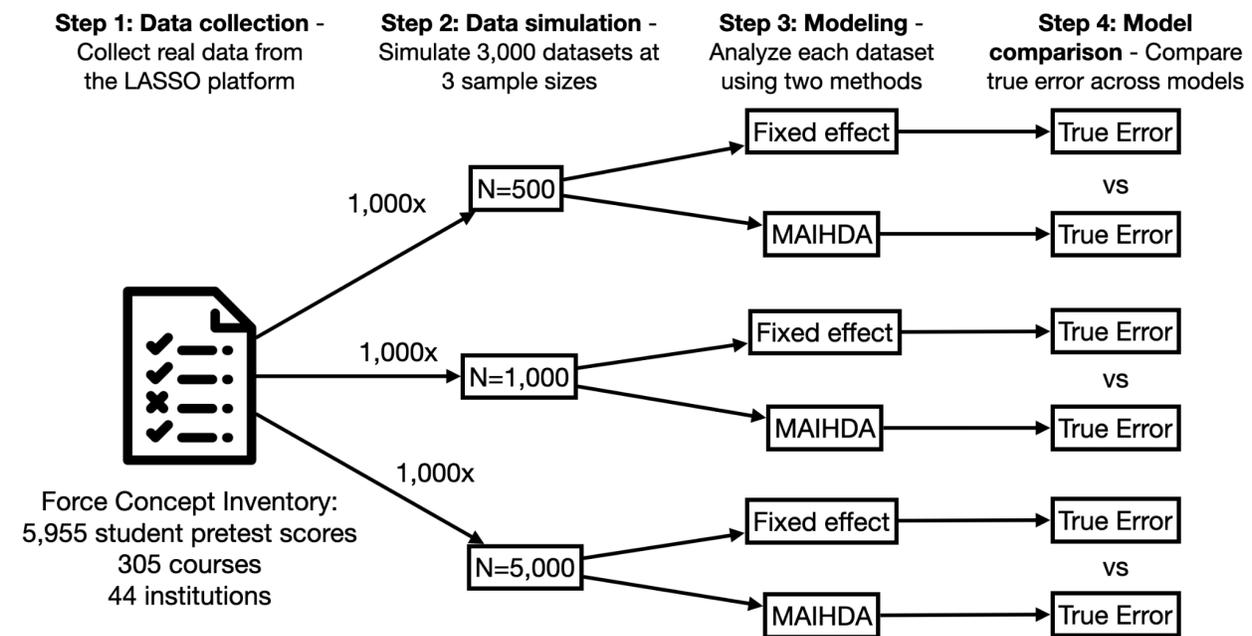

**Figure 2.** The five steps of our analysis. Step 1 used the Force Concept Inventory data to create a true model of test scores across 20 intersectional identities. Step 2 simulated data for the true model 1,000 times for each of the three sample sizes. Step 3 modeled the data using both fixed effect and MAIHDA models. Step 4 compared the true error (predicted - true score) across the fixed effect and MAIHDA models.

**Step 1: Data collection**

To ensure that our simulated analysis mirrored a realistic scenario in education research, we created simulated data based on real-world data collected through the Learning About STEM Student Outcomes (LASSO) platform (Van Dusen 2018). LASSO is an online assessment platform that hosts, administers, and analyzes research-based assessments for STEM instructors. LASSO provides instructors with reports on their student's progress on core science course content, then anonymizes responses from students who consent to share their data and makes it available to researchers. While the data in this study does not fully represent all institutional contexts, the LASSO database is more representative than the data typically used in existing studies of science students in higher education (Nissen et al. 2021). Appendix A details the institution types in the dataset as determined by the Carnegie classification of institutions of higher education (CCIHE) public 2021 database. Because LASSO was initially developed as part of the Learning Assistant Alliance, all of the instructors of the courses in this dataset reported engaging their students in collaborative learning. Thus, while there was institutional variation across courses, the courses represented in the sample featured less traditional lecture-style instruction than is typical in higher education physics courses.

LASSO provided the social identities and pretest scores from 5,955 students in 171 courses at 40 institutions on the Force Concept Inventory (FCI). Hestenes et al. (1992) developed the FCI to probe student understanding of Newtonian forces to assess the effectiveness of physics instruction. Researchers have applied many different quantitative methods to data from the FCI (e.g., Eaton and Willoughby 2020). One strand of this quantitative research focuses on the fairness of the FCI and its ability to produce unbiased data across different strata. While Morley et al. (2023) found that the FCI was measurement invariant across the intersection of 10 racial and gender strata, evidence indicates that several items on the FCI function differently for men (Traxler et al. 2018) and White men in particular (Buncher et al. 2021) than for individuals identifying with other gender or racial strata. Nonetheless, researchers have often used the FCI to investigate the effectiveness of instructional techniques (Han et al. 2015; Xiao et al. 2020; Bruun and Brewe 2013; Caballero et al. 2012; Nissen et al. 2022) and equity in courses (Good et al. 2019; Brewe et al. 2009; Van Dusen and Nissen 2020).

Society's educational debts owed to marginalized groups were calculated by subtracting the mean score for a strata from those of CG White men.

**Step 2: Data Simulation**

To span the sample sizes typically seen in equity studies (Van Dusen and Nissen 2022), we simulated datasets with total sample sizes of 500, 1,000, and 5,000 students. We simulated 1,000 datasets at each sample size for a total of 3,000 simulated datasets. Each student in the dataset was nested within a course with 50 students. We set the standard deviation in course average scores to be 10 points; the range of performance possible on the FCI is 0-100. We included this variance as it matches real-world data and is a common feature in educational contexts (Van Dusen and Nissen 2019; 2020; Sun and Pan 2014; Condon et al. 2016). We did not create a more nuanced student-classroom structure (e.g., having men more represented in higher or lower-performing contexts) as it was beyond the scope of this study.

Forty-four of the 1,000 generated datasets with a total sample size of 500 had produced at least one strata with no data. As these could not be analyzed with either the fixed effects or MAIHDA models, we excluded them from the analysis. Removing the datasets with strata sample sizes of zero resulted in a total of 2,956 simulated datasets for analysis.

Within each simulated dataset, we included strata variables for five racial groups (i.e., Asian (14%), Black (6%), Hispanic (7%), White (63%), and White Hispanic (10%)), two gender groups (i.e., women (36%) and men (64%)), and two college-generation groups (i.e., first-generation (FG; 36%) or continuing-generation (CG; 64%)), creating 20 intersectional social strata combinations in total. Based on the LASSO data, we set each strata's representation rates, true values, and standard deviations. The true score and proportional representation for each strata and the student-level standard deviations can be seen in Table 3. The simulation set the proportional representation of each strata in the dataset, but how they intersected varied across each simulation. For example, we set 6% of the students in each simulation as Black, 36% as women, and 36% as FG. We assigned each axis of social identifiers (i.e., race, gender, and FG/CG designation) independently, meaning that we did not ensure that 36% of each racial group were women. So, while FG Black women made up 0.76% (6% x 36% x 36%) of the data on average, in any given simulation, it could range from 0% (e.g., if none of the Black students were women and FG) to 6% (if all of the Black students were women and FG). By allowing the proportion of students in each strata to vary across simulations, we were able to examine the accuracy of predicted outcomes across a broader range of strata sample sizes.

*Student score simulation equation*

$$\begin{aligned}Score_{ij} = {} & 47 - 3 * FG_{ij} - 10 * women_{ij} - 6 * Black_{ij} - 10 * Hispanic_{ij} + 2 * White_{ij} + 3 \\ & * FG_{ij} * women_{ij} + 1 * White_{ij} * Hispanic_{ij} - 1 * FG_{ij} * Black_{ij} + 2 * FG_{ij} \\ & * Hispanic_{ij} + 2 * FG_{ij} * White_{ij} - 2 * women_{ij} * Black_{ij} + 1 * women_{ij} \\ & * Hispanic_{ij} - 4 * women_{ij} * White_{ij} - 1 * FG_{ij} * White_{ij} * Hispanic_{ij} + 5 \\ & * women_{ij} * White_{ij} * Hispanic_{ij} + 4 * FG_{ij} * women_{ij} * Black_{ij} + 1 * FG_{ij} \\ & * women_{ij} * White_{ij} - 1 * FG_{ij} * women_{ij} * White_{ij} * Hispanic_{ij} \\ & + u\_course_j + r_i \end{aligned}$$

$u\_course_j \sim N(0,10)$ (level 2)
$r_i \sim N(0,20)$ (level 1)

*(1)*

Similarly, our simulation randomly assigned students to courses. For instance, while 36% of the population were women, within any given course, the share of women could range from 0% to 100%. We set the standard deviation for scores within a strata to be 20 points.

**Step 3: Modeling**

We analyzed each dataset using a fixed effects model and a MAIHDA model. Our fixed effects intersectional model (Figure 3) was a frequentist multilevel model that nested students (level 1) within courses (level 2). The fixed effects intersectional model included an interaction term for each combination of race, gender, and college-generation terms.

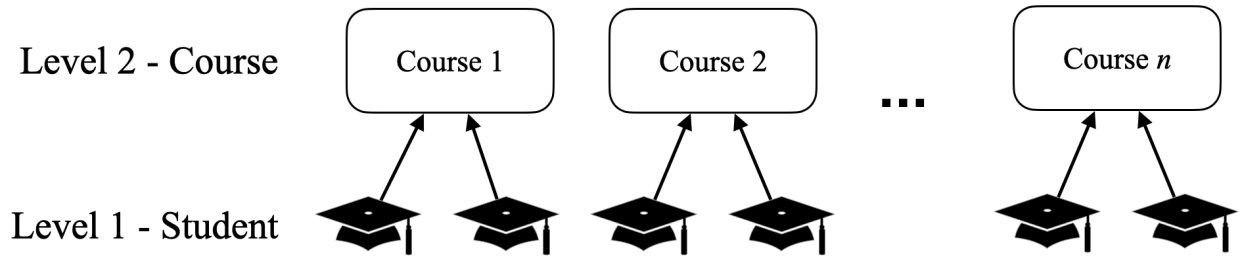

**Figure 3.** The multilevel structure of our fixed effects model with students (level 1) nested within courses (level 2).

*Fixed effects model*

$$\begin{aligned}
Score_{ij} = {} & \beta_0 + \beta_1 Black_{ij} + \beta_2 Hispanic_{ij} + \beta_3 White_{ij} + \beta_4 Hispanic_{ij} \times White_{ij} \\
& + Woman_{ij} \times (\beta_5 + \beta_6 Black_{ij} + \beta_7 Hispanic_{ij} + \beta_8 White_{ij} \\
& + \beta_9 Hispanic_{ij} \times White_{ij}) \\
& + FG_{ij} \times (\beta_{10} + \beta_{11} Black_{ij} + \beta_{12} Hispanic_{ij} + \beta_{13} White_{ij} \\
& + \beta_{14} Hispanic_{ij} \times White_{ij}) \\
& + FG_{ij} \times Woman_{ij} \times (\beta_{15} + \beta_{16} Black_{ij} + \beta_{17} Hispanic_{ij} + \beta_{18} White_{ij} \\
& + \beta_{19} Hispanic_{ij} \times White_{ij}) + u\_course_j + e_{ij}
\end{aligned}$$

$u\_course_j \sim N(0, \sigma^2_{u\_course})$ (level 2)
$e_{ij} \sim N(0, \sigma^2_e)$ (level 1)

*(2)*

In our equations, we follow the notation of Fielding and Goldstein (2006) for multilevel models and cross-classified multilevel model equations. The subscripts, for example, $Score_{ij}$, refer to the score for the *i*th student in the *j*th course. The $\beta_0$ term represents the score for continuing-generation Asian men. The $\beta_1$ term represents the shift from Asian to Black student scores. The $e_{ij}$ term represents the student-level error for a specific score, is the difference between the predicted and actual values, and is analogous to the ε term in standard linear regressions. The $u\_course_j$ term represents the course-level error for each course and allows the intercept to vary across each course. None of the variables were centered.

Our MAIHDA model (Figure 4) was a Bayesian cross-classified multilevel model that nested students (level 1) within courses (level 2a) and strata (level 2b).

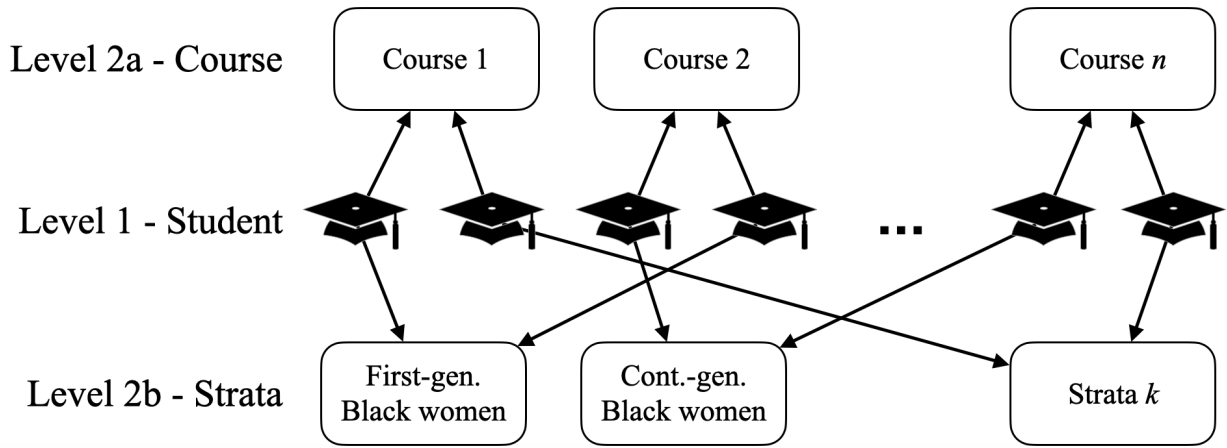

**Figure 4.** The multilevel structure of our cross-classified MAIHDA model with students (level 1) nested within courses (level 2a) and strata (level 2b). The strata lie at the intersections of social identity variables included in the model, and the model includes an error term for each strata.

*MAIHDA model*

$$Score_{i(j_1,j_2)} = \beta_0 + \beta_1 Black_{(j_1,j_2)} + \beta_2 Hispanic_{(j_1,j_2)} + \beta_3 White_{(j_1,j_2)} + \beta_4 Woman_{(j_1,j_2)} \\ + \beta_5 FG_{(j_1,j_2)} + u\_course_{j_1} + u\_strata_{j_2} + e_{i(j_1,j_2)}$$

$u\_course_{j_1} \sim N(0, \sigma^2_{u\_course})$ (Level 2a)
$u\_strata_{j_2} \sim N(0, \sigma^2_{u\_strata})$ (Level 2b)
$e_{i(j_1,j_2)} \sim N(0, \sigma^2_e)$ (Level 1)

*(3)*

The subscripts, for example, $Score_{i(j_1,j_2)}$, refer to the score for the *i*th student in the $j_1$th course and the $j_2$th strata. $j_1$ runs from 1 to *n*, where *n* represents the total number of courses. $j_2$ runs from 1 to *k*, where *k* represents the total number of strata. The model identified *n* and *k* as the number of unique values in the course and strata variables. In this study, *n* varied across the sample sizes of 500 (*n* = 10), 1,000 (*n* = 20), and 5,000 (*n* = 100). *k* was 20 due to the number of unique combinations of strata variables (5 race x 2 gender x 2 college generation designations). In addition to including student- and course-level error terms ($e_{i(j_1,j_2)}$ and $u\_course_{j_1}$) in the fixed effect model, the MAIHDA equation includes a strata error term ($u\_strata_{j_2}$). Researchers can let the algorithms automatically identify the unique strata (e.g., Evans et al. 2018) or set the strata that are included (e.g., strata with samples of 20 or more). The two practices are equivalent for our simulated data. While we realize that the difference between the two practices may materialize with real-world data where many intersections have few participants (e.g., Evans et al. 2018), that question is beyond the scope of this paper. None of the variables were centered.

To understand the structure of the variance of our data we ran an unconditional cross-classified model with no fixed effects. We used this model to calculate the variance partition coefficient (VPC) and the Proportional change in variance (PCV). The VPC measures the variance attributed to a level in a multilevel model and is sometimes called the intraclass correlation coefficient. In most MAIHDA models, the VPC score for strata is often <5% and usually <10% (Evans et al. 2020). The PCV measures the total between-stratum variation explained by including additive main effects.

*Unconditional cross-classified multilevel model*

$$Score_{i(j_1,j_2)} = \beta_0 + u\_course_{j_1} + u\_strata_{j_2} + e_{i(j_1,j_2)}$$

$u\_course_{j_1} \sim N(0, \sigma^2_{u\_course})$ (Level 2a)
$u\_strata_{j_2} \sim N(0, \sigma^2_{u\_strata})$ (Level 2b)
$e_{i(j_1,j_2)} \sim N(0, \sigma^2_e)$ (Level 1)

*(4)*

*Table 2.* The variance, VPC, and PCV for the real-world and simulated datasets.

| | $\sigma_e^2$ | $\sigma_{u\_course}^2$ | $\sigma_{u\_strata}^2$ | $VPC_{course}$ | $VPC_{strata}$ | $PCV_{strata}$ |
|---|---|---|---|---|---|---|
| **Real-world data** | | | | | | |
| Unconditional CCMM Model | 423.4 | 17.4 | 53.8 | 3.6% | 11.2% | - |
| MAIHDA Model | 415.8 | 17.7 | 9.2 | 4.1% | 2.5% | 78.0% |
| **Simulated data** | | | | | | |
| Unconditional CCMM Model | 399.9 | 109.7 | 45.7 | 20% | 9% | - |
| MAIHDA Model | 399.6 | 109.6 | 6.5 | 21% | 2% | 79.6% |

$\sigma_e^2$ = Student-level variance.
$\sigma_{u\_course}^2$ = Course-level variance.
$\sigma_{u\_strata}^2$ = Strata-level variance
$VPC_{course}$ = Course-level variance partition coefficient, converted to a percent.
$VPC_{strata}$ = Strata-level variance partition coefficient, converted to a percent.
$PCV_{strata}$ = Strata-level proportional change in variance, converted to a percent. The PCV represents the percent of the total between-strata variation that is explained by inclusion of primary terms.

    Comparing the variances from the real-world data with our simulated data shows that the share of the variance accounted for by strata (VPC) is very similar in both datasets (Table 2). Including the fixed effects terms in the MAIHDA model also accounts for similar shares of the variance due to strata (PCV) for the real-world and simulated data sets. The one place where the two datasets diverge is the share of the variance accounted for by the courses. We reran our simulation with a smaller standard deviation in course scores (4.5 points). This made the $VPC_{course}$ values match the real-world data, but did not meaningfully impact any of the group predicted outcomes. This shows that our results are robust across a range of variance at the course-level. It should be noted, however, that our course-level variance does not have any structure (i.e., there is no correlation between course and strata-level variances).

    Our unconditional model showed that the $VPC_{strata}$ was 9%, within the typical range reported (Evans et al. 2020). The $PCV_{strata}$ in the MAIHDA model was 79.6%, which means that the primary terms in our MAIHDA model accounted for 79.6% of the variance in the strata level. It also means that 20.4% of the variance between strata can be attributed to intersectional effects.

**Step 4: Model comparison**
    To compare the precision and accuracy of MAIHDA and fixed effect models and inform minimum sample sizes for robust estimation, we calculated the true error in the prediction for each of the 20 strata. The true error is the difference between the predicted and true scores (estimated score minus true score). We compared their mean absolute true errors to determine

which modeling technique yielded more accurate predictions. To determine which modeling technique would allow for the modeling of smaller strata sample sizes, we examined the mean absolute true error for strata with sample sizes of 20 or less.

## Findings

*RQ 1. What are typical strata representation rates and outcomes on physics research-based assessments?*

LASSO provided data from 5,955 students in 171 courses at 40 institutions on the Force Concept Inventory (FCI). The data included 20 intersectional social strata (Table 3). The share of the sample for a strata ranged from 26% (CG White men) to 1% (CG Black women; FG Black men; FG Black women; FG Hispanic women; and FG White Hispanic women). The mean score on the assessment for a strata ranged from 49 points (CG White men) to 28 points (CG Hispanic women). Society's educational debts ranged from 21 points (CG Hispanic women) to 2 points (CG Asian men).

*Table 3.* The share of the total sample, true score, and student-level standard deviation for each intersectional social strata as set by the simulation.

| Intersectional Social Strata | Share of sample | Score (points) | Society's educational debt (points) |
|---|---|---|---|
| CG Asian men | 6% | 47 | 2 |
| CG Asian women | 3% | 37 | 12 |
| CG Black men | 2% | 41 | 8 |
| CG Black women | 1% | 29 | 20 |
| CG Hispanic men | 3% | 37 | 12 |
| CG Hispanic women | 2% | 28 | 21 |
| CG White Hispanic men | 4% | 40 | 9 |
| CG White Hispanic women | 2% | 32 | 17 |
| CG White men | 26% | 49 | - |
| CG White women | 15% | 35 | 14 |
| FG Asian men | 3% | 44 | 5 |
| FG Asian women | 2% | 37 | 12 |
| FG Black men | 1% | 37 | 12 |
| FG Black women | 1% | 32 | 17 |
| FG Hispanic men | 2% | 36 | 13 |
| FG Hispanic women | 1% | 30 | 19 |
| FG White Hispanic men | 2% | 40 | 9 |
| FG White Hispanic women | 1% | 35 | 14 |
| FG White men | 15% | 48 | 1 |
| FG White women | 8% | 38 | 11 |

*RQ 2. To what extent do MAIHDA models create more accurate predicted* strata *outcomes than fixed effects models?*

We compared their true errors to determine the accuracy of the two modeling methods. Table 4 shows the mean absolute true error disaggregated by model, strata, and total sample size. The MAIHDA model had smaller mean absolute true errors in 56 of the 60 cases (i.e., 20 strata in each of the three total sample sizes). In the four exceptions in which the MAIHDA model was less accurate on average, the mean absolute true errors were only marginally larger, ranging from 0.0-0.4 points. As the total sample sizes increased, the amount by which MAIHDA outperformed the fixed effects model decreased. MAIHDA's mean absolute true error was 31% (1.7 points) smaller than the fixed effects model when the total sample size was 500 but only 13% (0.2 points) smaller when the sample size was 5,000.

*Table 4.* Each strata's mean absolute true error across the three total sample sizes. The delta mean absolute true error is the MAIHDA mean absolute true error minus the fixed effect mean absolute true error. Negative delta values indicate that the MAIHDA model generally produces less error.

| Intersectional Social Strata | True score | Total sample size = 500 | | | | Total sample size = 1,000 | | | | Total sample size = 5,000 | | | |
|---|---|---|---|---|---|---|---|---|---|---|---|---|---|
| | | Mean N | Mean absolute true error | | | Mean N | Mean absolute true error | | | Mean N | Mean absolute true error | | |
| | | | Fixed Effect | MAIHDA | Delta | | Fixed Effect | MAIHDA | Delta | | Fixed Effect | MAIHDA | Delta |
| CG Asian men | 47 | 28.3 | 4.0 | 3.4 | -0.6 | 56.8 | 2.8 | 2.5 | -0.3 | 285.9 | 1.2 | 1.2 | 0.0 |
| CG Asian women | 37 | 15.9 | 4.7 | 3.6 | -1.1 | 32.3 | 3.5 | 2.6 | -0.9 | 160.8 | 1.5 | 1.3 | -0.2 |
| CG Black men | 41 | 11.8 | 5.4 | 4.2 | -1.2 | 23.6 | 3.8 | 3.0 | -0.8 | 117.4 | 1.7 | 1.6 | -0.1 |
| CG Black women | 29 | 6.7 | 6.8 | 4.3 | -2.5 | 13.1 | 4.8 | 3.0 | -1.8 | 65.7 | 2.2 | 1.6 | -0.6 |
| CG Hispanic men | 37 | 15.1 | 5.0 | 3.7 | -1.3 | 30.2 | 3.6 | 2.7 | -0.9 | 151.3 | 1.5 | 1.3 | -0.2 |
| CG Hispanic women | 28 | 8.5 | 6.2 | 3.5 | -2.7 | 17.1 | 4.7 | 2.8 | -1.9 | 84.5 | 1.9 | 1.4 | -0.5 |
| CG White Hispanic men | 40 | 20.2 | 4.7 | 3.5 | -1.2 | 40.5 | 3.1 | 2.5 | -0.6 | 200.9 | 1.4 | 1.2 | -0.2 |
| CG White Hispanic women | 32 | 11.4 | 5.5 | 3.9 | -1.6 | 22.8 | 3.9 | 2.8 | -1.1 | 113.5 | 1.7 | 1.4 | -0.3 |
| CG White men | 49 | 129.0 | 2.9 | 2.9 | 0.0 | 258.2 | 2.1 | 2.1 | 0.0 | 1292.5 | 0.9 | 0.9 | 0.0 |
| CG White women | 35 | 72.9 | 3.1 | 3.1 | 0.0 | 145.1 | 2.2 | 2.4 | 0.2 | 726.9 | 1.0 | 1.1 | 0.1 |
| FG Asian men | 44 | 16.2 | 4.8 | 3.7 | -1.1 | 31.9 | 3.4 | 2.7 | -0.7 | 160.8 | 1.6 | 1.4 | -0.2 |
| FG Asian women | 37 | 9.1 | 5.7 | 3.7 | -2.0 | 18.3 | 4.3 | 2.8 | -1.5 | 90.4 | 2.0 | 1.5 | -0.5 |
| FG Black men | 37 | 6.7 | 7.3 | 4.6 | -2.7 | 13.3 | 4.7 | 3.5 | -1.2 | 66.1 | 2.1 | 2.0 | -0.1 |
| FG Black women | 32 | 3.7 | 9.3 | 4.9 | -4.4 | 7.4 | 6.6 | 3.4 | -3.2 | 37.1 | 2.8 | 2.1 | -0.7 |
| FG Hispanic men | 36 | 8.5 | 6.1 | 3.9 | -2.2 | 16.8 | 4.5 | 3.3 | -1.2 | 85.1 | 1.9 | 1.7 | -0.2 |
| FG Hispanic women | 30 | 4.8 | 8.7 | 4.0 | -4.7 | 9.6 | 5.6 | 2.9 | -2.7 | 47.6 | 2.4 | 1.7 | -0.7 |
| FG White Hispanic men | 40 | 11.4 | 5.7 | 3.6 | -2.1 | 22.7 | 3.8 | 2.6 | -1.2 | 113.2 | 1.7 | 1.3 | -0.4 |
| FG White Hispanic women | 35 | 6.4 | 7.3 | 5.2 | -2.1 | 12.7 | 4.9 | 4.1 | -0.8 | 63.8 | 2.3 | 2.7 | 0.4 |
| FG White men | 48 | 72.5 | 3.2 | 3.0 | -0.2 | 146.0 | 2.3 | 2.2 | -0.1 | 726.5 | 1.0 | 1.0 | 0.0 |
| FG White women | 38 | 40.9 | 3.5 | 3.1 | -0.4 | 81.8 | 2.6 | 2.3 | -0.3 | 409.9 | 1.2 | 1.1 | -0.1 |
| **Total** | | **25.0** | **5.5** | **3.8** | **-1.7** | **50.0** | **3.9** | **2.8** | **-1.1** | **250.0** | **1.7** | **1.5** | **-0.2** |

The largest mean absolute true error for the fixed effect model was for FG Black women (9.3 points), who also had the smallest mean N (3.7). In the MAIHDA model, the mean absolute true error was reduced to 4.9 points. Given that the standard deviation for student scores was 20 points, the shift from fixed effect models to MAIHDA models decreased the mean absolute true error from 0.47 SD to 0.25 SD.

To understand how strata sample sizes impacted the accuracy of the models, we examined the true error for each of the strata predictions for each model. Figure 5 shows the true error for all the predictions versus each model's strata sample size. Both models were similarly accurate when they had larger total and strata sample sizes. The magnitude of the true errors increased for both models with smaller total sample sizes and strata sample sizes. Where the models differed, however, was that the magnitude of the true error increased more quickly in the fixed effects model as the strata sample sizes decreased. The improvement in performance for MAIHDA over fixed effects models was most apparent for the strata with the smallest sample size (i.e., FG Black women). The reduction in the mean absolute true error for FG Black women when using MAIHDA ranged from 47% with a total sample size of 500 to 25% with a total sample size of 5,000.

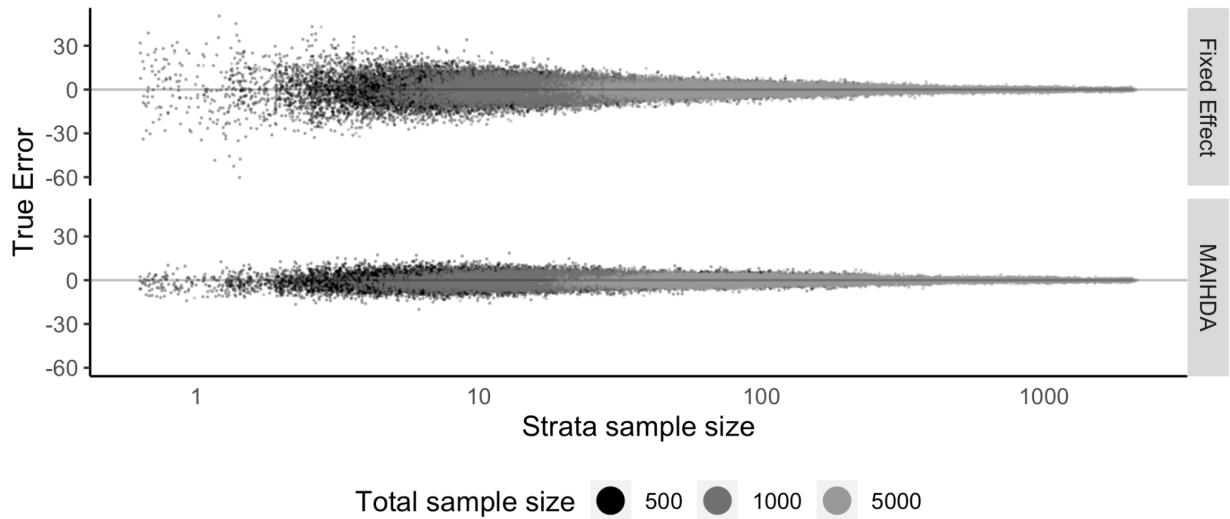

**Figure 5**. A logarithmic scatter plot of the true error for each predicted strata outcome versus the strata sample size for each model. The shade of the dots represents the total sample sizes. Predicted values were jittered to enhance visibility.

*RQ 3. To what extent do MAIHDA models allow the modeling of smaller strata sample sizes?*

To understand the differences in each model's ability to handle smaller strata sample sizes, we examined the absolute mean and mean of the true error for each strata sample size (Table 5). Figure 6 shows the mean absolute true error values for strata sample sizes ranging from 1-20 across total sample sizes and modeling techniques. We limited the figure to this range, as Simmons and colleagues (2016) argue, using 20 as a minimum strata sample size. Figure 6 includes a dashed line as a comparison tool indicating the mean absolute true error for the fixed effects model when the strata sample size is 20 and the total sample size is 500 (4.4). For the MAIHDA model, the mean absolute true error did not reach 4.4 until the strata sample sizes were reduced to five or fewer.

*Table 5.* The mean and absolute mean of the true error values for each strata sample size disaggregated by model type and total sample size.

| Strata sample size | Total sample size = 500 | | | | Total sample size = 1,000 | | | |
|---|---|---|---|---|---|---|---|---|
| | Traditional | | MAIHDA | | Traditional | | MAIHDA | |
| | Absolute Mean True Error | Mean True Error | Absolute Mean True Error | Mean True Error | Absolute Mean True Error | Mean True Error | Absolute Mean True Error | Mean True Error |
| 1 | 15.68 | -0.68 | 5.29 | -2.93 | 13.17 | 2.73 | 5.56 | -3.58 |
| 2 | 11.16 | 0.59 | 4.70 | -1.77 | 11.26 | 2.69 | 2.31 | -0.53 |
| 3 | 9.54 | -0.41 | 4.82 | -1.35 | 8.22 | -1.63 | 3.09 | -1.42 |
| 4 | 8.59 | -0.16 | 4.62 | -1.08 | 8.69 | 0.28 | 3.59 | -2.34 |
| 5 | 7.55 | -0.05 | 4.41 | -0.73 | 8.27 | 1.50 | 3.75 | -1.34 |
| 6 | 7.03 | -0.43 | 4.24 | -0.70 | 6.16 | -0.26 | 3.40 | -1.88 |
| 7 | 6.24 | -0.16 | 4.07 | -0.37 | 6.27 | -0.29 | 3.50 | -1.70 |
| 8 | 6.16 | -0.25 | 4.16 | -0.43 | 5.98 | -0.01 | 3.28 | -1.03 |
| 9 | 5.87 | -0.14 | 4.01 | -0.26 | 5.43 | 0.11 | 3.36 | -1.06 |
| 10 | 5.46 | -0.05 | 3.84 | -0.18 | 5.13 | -0.10 | 3.33 | -0.49 |
| 11 | 5.51 | -0.35 | 3.93 | -0.51 | 5.33 | 0.67 | 3.56 | -0.23 |
| 12 | 5.34 | 0.02 | 3.84 | -0.03 | 5.07 | 0.04 | 3.44 | -0.18 |
| 13 | 4.75 | 0.24 | 3.60 | 0.17 | 4.81 | -0.21 | 3.33 | -0.34 |
| 14 | 4.82 | -0.25 | 3.63 | -0.16 | 4.58 | -0.05 | 3.19 | 0.02 |
| 15 | 4.73 | -0.38 | 3.62 | -0.14 | 4.56 | -0.34 | 3.08 | -0.06 |
| 16 | 4.85 | -0.21 | 3.66 | -0.06 | 4.36 | -0.09 | 2.99 | -0.08 |
| 17 | 4.94 | -0.25 | 3.73 | -0.06 | 4.21 | 0.16 | 3.02 | 0.42 |
| 18 | 4.76 | 0.06 | 3.66 | 0.14 | 4.10 | 0.67 | 2.88 | 0.49 |
| 19 | 5.04 | 0.48 | 3.93 | 0.59 | 4.21 | 0.47 | 2.91 | -0.09 |
| 20 | 4.39 | -0.19 | 3.40 | -0.01 | 4.14 | 0.31 | 2.80 | 0.15 |

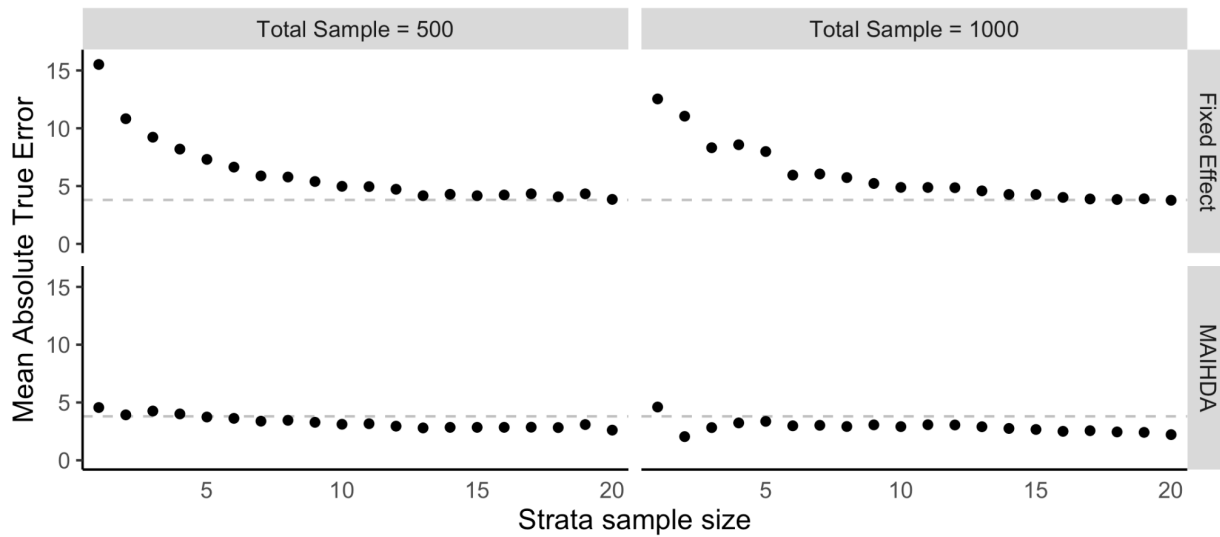

**Figure 6.** A plot of the mean absolute true error values by strata sample size across total sample size and modeling techniques.

For models with very small strata sample sizes, shrinkage will decrease the size of the strata's variance term. This smaller variance term leads to the strata's predicted outcomes being closer to the prediction produced by only adding the model's primary coefficients for the strata. In our MAIHDA models, when strata sample sizes fell below ten individuals, we began to have a small but consistent negative bias in the mean true error scores (see Table 5). The variance caused this negative bias for the least represented strata (first-generation Black women) to be positive on average. In other words, the additive effects of the coefficients for the intercept, first-generation, Black, and women predicted scores lower than the true score for first-generation Black women. Having fewer data for that strata regressed their predicted score closer to the additive score from the model and created a negative bias.

# Discussion

A central goal of intersectional modeling is to disaggregate intersectional social strata outcomes as much as possible to represent the diversity of lived experiences. MAIHDA was developed to support further disaggregation of data across strata than is allowable by fixed effects regression models. To identify whether MAIHDA is superior to fixed effects models at predicting outcomes for small strata, we compared the accuracy of both modeling techniques on simulated datasets that mirrored real-world physics student outcomes.

Our findings indicated that meaningful inequities exist in physics student scores on the Force Concept Inventory and that MAIHDA created more accurate models of these inequities than the fixed effects models did (Table 4 and Figure 5). When using the MAIHDA model rather than the fixed effects model, we saw the most significant improvements in accuracy with predictions made for small total sample sizes and strata with smaller sample sizes. These improvements directly support the goal of using MAIDHA to create models with *smaller* strata, which coincidentally allows for *more* strata to reflect the scope of intersectional experiences within a population.

# Conclusion

The accuracy improvements observed by employing MAIHDA over fixed effects models are strong evidence that MAIHDA is achieving its creator's goal of improving the quality of predicted outcomes across intersectional social strata. These results also suggest that researchers may be able to disaggregate their models further, increasing the number of distinct strata represented by accommodating smaller strata sizes. Our examination of the mean absolute true error for small strata sample sizes bore out this idea (Table 5 and Figure 6). One recommended value for minimum strata sample sizes with fixed effects models is 20 (Simmons, Nelson, and Simmonsohn 2016). For our fixed effects models, the mean absolute true error was very similar for total sample sizes of 500 and 1,000 when the strata sample sizes were 20 or less. The MAIHDA model maintained an accuracy equal to or better than those observed in the fixed effects model until the sample sizes were five or fewer. These smaller standard error values indicated that researchers could model strata sample sizes below 20 without meaningful compromises in accuracy when using MAIHDA. We refrain from making specific minimum strata sample size recommendations as the context of each research project will determine how much potential error a researcher is willing to entertain to increase the number of strata a model includes.

Beyond being a more effective methodological tool, MAIHDA can change how intersectionality is theorized and reified in quantitative research. Introducing tools to a system can transform cognitive tasks (Hutchins 1995a; 1995b) and lead to novel outcomes (Roth and Lee 2007). It can be challenging to imagine the things that we have never seen done before. By allowing models to include strata with a sample size of one, MAIHDA can include all participants as they self-identify in an analysis without aggregating groups or engaging in data erasure. Opening up this methodological possibility will require researchers to re-envision intersectional theory and its place in quantitative research.

While our findings provide strong evidence of the superiority of using MAIHDA when modeling intersectional outcomes, it is only appropriate for some scenarios. Specifically, MAIHDA requires a sufficient number of strata to offer improved predictions. Evans and colleagues (2018) recommend having at least 20 strata in a model before using them as a level to nest students in to ensure the model has sufficient level-2 random effects. Researchers (e.g., Silva and Evans 2020; Evans 2019) have begun to explore the use of MAIHDA models with fewer strata, but the impact of having fewer strata in a model has yet to be fully known. While there are examples of education equity research with models that include 20 or more strata (e.g., Jang 2018; Nissen and Van Dusen under review; Van Dusen and Nissen under review), many have not (Van Dusen and Nissen 2022). With the increased use of novel modeling methods, large-scale datasets (e.g., LASSO), and intersectional perspectives in quantitative research (Wofford and Winkler 2022), the trend may be shifting, however, to models that offer more nuanced depictions of marginalized students' outcomes through the inclusion of more strata.

While researchers are investigating the implications of using MAIHDA to create intersectional models, the statistical technique of blending fixed and random effects to create predicted outcomes is applicable across many areas of quantitative research. Just as Jones and colleagues (2016) used an analogous technique to MAIHDA to investigate voting rates, judiciously replacing some fixed effects with random effects can improve model accuracy. When using MAIHDA, researchers have primarily examined replacing interaction terms with random effects (Keller et al. 2023), but the most efficacious replacement of fixed effects will be contextually dependent. Additional research is needed to identify the best method for selecting fixed effect terms in MAIHDA or analogous non-intersectional models.

# Future research

Developing and running the MAIHDA and the fixed effect models took similar effort with our simulated dataset. However, one feature of our simulated data that differed from many educational datasets is that it lacked any missing data. In these situations, multiple imputations are often recommended to maximize statistical power and limit the introduction of bias (Rubin 1996; Nissen et al. 2019). Examining multiple imputed datasets can complicate analyses. Future research should compare the practicality and efficacy of using MAIHDA and fixed effect models with a real-world dataset that includes missing values.

While we did not examine it in this study, the Bayesian nature of MAIHDA models means their accuracy could be improved through informed priors. Informed priors allow researchers to directly include findings from prior research into their models, similar to what is done in a meta-analysis. While an investigation may not have the statistical power to normally include a particular strata in a model, if it can draw on findings across other investigations, it can create reasonably accurate predictions for even smaller strata. Determining the impact of using informed priors with MAIHDA models will require future research.

This analysis did not examine how to account for interaction effects between strata and other factors. For example, when using a fixed effect model to determine the differential impact of a classroom intervention across strata, researchers can add fixed interaction terms between the intervention and intersectional social strata variables. In a MAIHDA model, however, the interaction effect is between a fixed variable (e.g., the intervention) and a random variable (e.g., intersectional social strata). It must be included as a random effect term. While modeling programs can run such models, determining the impact of moving these interaction terms from fixed to random will require future research.

This study examined continuous outcomes. While the efficacy we found is likely to exist for logistic models of categorical outcomes, a simulation study is warranted to test it empirically.

This study provided proof that MAIHDA can work with cross-classified multilevel models, but the student-course structures used in our model were reasonably simple. Further research is needed to examine the efficacy of MAIHDA in modeling student outcomes with more complex student, course, and institution structures (e.g., creating subsets of courses that match the makeup of highly, medium, and low-selectivity institutions). Future simulation studies could also examine the impact of creating cross-classified multilevel models with levels for course and strata versus simply including the strata level.

Equity-minded researchers are cautious about treating strata monolithically. While intersectional research that considers variables such as those introduced here somewhat addresses this concern, more work is needed to explore the potential for disaggregation of data according to more specific racial designations, such as regional Hispanic populations. In this work, we illustrate the potential of MAIHDA using demographic data likely to exist in many surveys that ask standard demographic questions to report on broad categories related to race, gender, and socioeconomic identification (or its proxies, e.g., college generation). More work is needed to illustrate the potential of MAIHDA–and quantitative intersectionality work generally–using more precise demographic data such as immigration generation, family country of origin, or country region.

# Research Ethics Statement

As this research does not constitute human subjects research, IRB approval was not necessary to carry out this work.

Van Dusen, Ben, Jayson Nissen, Robert M. Talbot, Hannah Huvard, and Mollee Shultz. "A QuantCrit investigation of society's educational debts due to racism and sexism in chemistry student learning." *Journal of Chemical Education* 99, no. 1 (2021): 25-34.

Vrieze, S. I. 2012. "Model selection and psychological theory: a discussion of the differences between the Akaike information criterion (AIC) and the Bayesian information criterion (BIC)". *Psychological methods*, 17(2), 228.

Wasserstein, R. L., and Lazar, N. A. 2016. "The ASA statement on p-values: context, process, and purpose." The American Statistician, 70(2), 129-133.

Willis, P. E. 1981. "Learning to labor: How working class kids get working class jobs." Columbia University Press.

Wofford, A. M., and Winkler, C. E. (2022). "Publication patterns of higher education research using quantitative criticism and QuantCrit perspectives." *Innovative Higher Education* 47 967-988.

Womack, Veronica Y., Letitia Onyango, Patricia B. Campbell, and Richard McGee. 2023. ""In the back of my mind": A Longitudinal Multiple Case Study Analysis of Successful Black Women Biomedical Graduate Students Navigating Gendered Racism." *CBE—Life Sciences Education* 22(3).

Xiao, Y., Xu, G., Han, J., Xiao, H., Xiong, J., and Bao, L. 2020. "Assessing the longitudinal measurement invariance of the Force Concept Inventory and the Conceptual Survey of Electricity and Magnetism." *Physical Review Physics Education Research, 16*(2), 020103.
31

# Appendix

*Appendix A*. Institution information from our dataset. Note: The subcategories do not add up the total because 2 institutions were not in the Carnegie classification of institutions of higher education (CCIHE) public 2021 database.

| Total | Type | | Size | | | Highest Degree | | | | Special Designators | |
|---|---|---|---|---|---|---|---|---|---|---|---|
| | Public | Private | Small | Medium | Large | AA | BA | MA | PhD | Hispanic-Serving Institution | Minority-Serving Institution |
| 40 | 25 | 13 | 5 | 15 | 18 | 4 | 5 | 14 | 15 | 11 | 11 |